\documentclass[11pt]{article}
\usepackage[a4paper,top=2.5cm,bottom=3cm,left=2.5cm,right=2.5cm]{geometry}

\usepackage{framed,multirow}
\usepackage{authblk}
\usepackage[dvipsnames]{xcolor}
\usepackage[T1]{fontenc}
\usepackage[utf8]{inputenc}
\usepackage{xcolor}
\usepackage{amsfonts}
\usepackage{pifont}
\usepackage{tikz}
\usepackage{longtable}
\usepackage{booktabs}
\usepackage{array}

\usepackage{hyphenat}

\usepackage{svg}
\usepackage{lscape}
\usepackage{graphicx}
\usepackage{longtable}
 \usepackage{rotating}
\usepackage{booktabs}
\usepackage{makecell}
\usepackage{orcidlink}

\usepackage[numbers]{natbib} 

\usepackage{adjustbox}
\usepackage{fancybox,framed}
\usepackage{hyperref}
\usepackage{comment}
\hypersetup{
    colorlinks=true,
    linkcolor=blue,
    filecolor=magenta,      
    urlcolor=blue,
    pdftitle={Overleaf Example},
    pdfpagemode=FullScreen
    }

\title{Exploring Adult Glioma through MRI: A Review of Publicly Available Datasets to Guide Efficient Image Analysis
} 

\author[2,1]{Meryem Abbad Andaloussi\orcidlink{0009-0003-9274-2086}}

\author[1]{Raphael Maser\orcidlink{0009-0006-8023-6292}}

\author[3]{Frank Hertel\orcidlink{0000-0002-2242-4045}}

\author[1]{François Lamoline\orcidlink{0000-0003-4289-2329}\textsuperscript{*}} 

\author[1]{Andreas Dominik Husch\orcidlink{0000-0001-9404-5127}\textsuperscript{*}}

\affil[1]{Imaging AI Group, Luxembourg Centre for Systems Biomedicine, University of Luxembourg, Belvaux, Luxembourg}
\affil[2]{Faculty of Science, Technology and Medicine, University of Luxembourg,  University of Luxembourg, Belvaux, Luxembourg}

\affil[3]{National Department of Neurosurgery, Centre Hospitalier de Luxembourg, Luxembourg, Luxembourg}
\affil[*]{These authors contributed equally to this work.}
\affil[ ]{Correspondence: andreas.husch@uni.lu}

\begin{document}
\maketitle
\begin{abstract}

Publicly available data is essential for the progress of medical image analysis, in particular for crafting machine learning models. Glioma is the most common group of primary brain tumors, and magnetic resonance imaging (MRI) is a widely used modality in their diagnosis and treatment. However, the availability and quality of public datasets for glioma MRI are not well known. In this review, we searched for public datasets for glioma MRI using Google Dataset Search, The Cancer Imaging Archive (TCIA), and Synapse. A total of 28 datasets published between 2005 and May 2024 were found, containing 62019 images from 5515 patients. We analyzed the characteristics of these datasets, such as the origin, size, format, annotation, and accessibility. Additionally, we examined the distribution of tumor types, grades, and stages among the datasets. 
The implications of the evolution of the WHO classification on tumors of the brain are discussed, in particular the 2021 update that significantly changed the definition of glioblastoma. 
Additionally, potential research questions that could be explored using these datasets were highlighted, such as tumor evolution through malignant transformation, MRI normalization, and tumor segmentation. Interestingly, only two datasets among the 28 studied reflect the current WHO classification. 
This review provides a comprehensive overview of the publicly available datasets for glioma MRI currently at our disposal, providing aid to medical image analysis researchers in their decision-making on efficient dataset choice.


\end{abstract}
\section{Introduction}
The study of glioma, some of the most prevalent brain tumor types, has been gaining interest due to advancements in imaging and modeling techniques. Despite their prevalence within the group of intracranial processes, they are overall still a rare disease, with an incidence rate of approximately 3 per 100,000 \cite{Pati22}, posing challenges in gathering extensive datasets for training reliable AI models. Various research groups have joined efforts to release a range of publicly available brain tumor datasets, each focusing on a distinct tumor type, study goal, and clinical setup. However, data scarcity remains a major challenge, in particular when considering the very diverse acquisition domains in clinical medical imaging, with different scanners, protocols, patient characteristics and budget constraints in globally diverse healthcare systems.

\subsection{Motivation}
The regular BraTS (Brain Tumor Segmentation) challenges \cite{brats2023}
and their respective datasets, have played a pivotal role in driving the development of brain tumor segmentation algorithms and have boosted the development of the field of medical image analysis overall. Over the last decade, it has helped ML researchers to develop, train, validate and refine their algorithms. 
While benchmark-based competitions have made significant contributions, there is a need to expand beyond these curated and annotated datasets from a single source: Such highly preprocessed data - in the example of BraTS even resampled to the same voxel size and aligned to a common space - may not adequately generalize to clinical MRI scans from different institutions and healthcare systems.  
This strong focus on highly pre-processed datasets with low variance could potentially introduce bias in models and result in poor out of domain generalizability.

More specifically for glioma tumors, the current available public datasets mostly provide MR imaging information. However, many of them lack complementary information such as histopathological confirmation of tumor type following the last World Health Organisation (WHO) classification or medical reports, which would lead to some errors in labelling or classification and be therefore not compliant with the current personalized medicine approach. 

\subsection{Scope and organization of the article}
The purpose of this study is to provide a comprehensive overview of publicly available  adult glioma MRI datasets and their different features to medical image analysis researchers, aiding them in more efficient method development. We evaluate 28 different adult glioma datasets between 2005 and 2023, presenting their properties and application scopes. Among the datasets, we show the complex BraTS inclusions. We present the main features of each dataset such as patients and images number, MRI modalities, tumor types, grades and corresponding WHO classification.

To the best of our knowledge, this is the first attempt to provide a comprehensive and comparative list of public adult glioma MRI datasets.
In 2022, Yearley et al. \cite{yearley2022current} provided a more general overview of various glioma data registries, including clinical and molecular data resources available for glioma research. However, the authors did not delve into the specific details of MRI imaging studies related to adult glioma and consequently the implications of MRI features in diagnosis, monitoring and treating of adult gliomas are not discussed. In contrast, the present work has the needs of medical image analysis in focus.\\ 

The article is organized as follows. First, the search methodology leading to selection of datasets is presented in section \ref{sec:search}. The search results and the datasets characteristics are presented in section \ref{sec:result}. Additionally, in Section \ref{sec:discu}, we delve into the practical applications and challenges of these datasets in addressing various potential research questions. Finally, we conclude with perspectives for future works.

\section{Search methodology}
\label{sec:search}
\begin{figure*}[ht]
    \centering
    \includegraphics[width=0.85\linewidth]{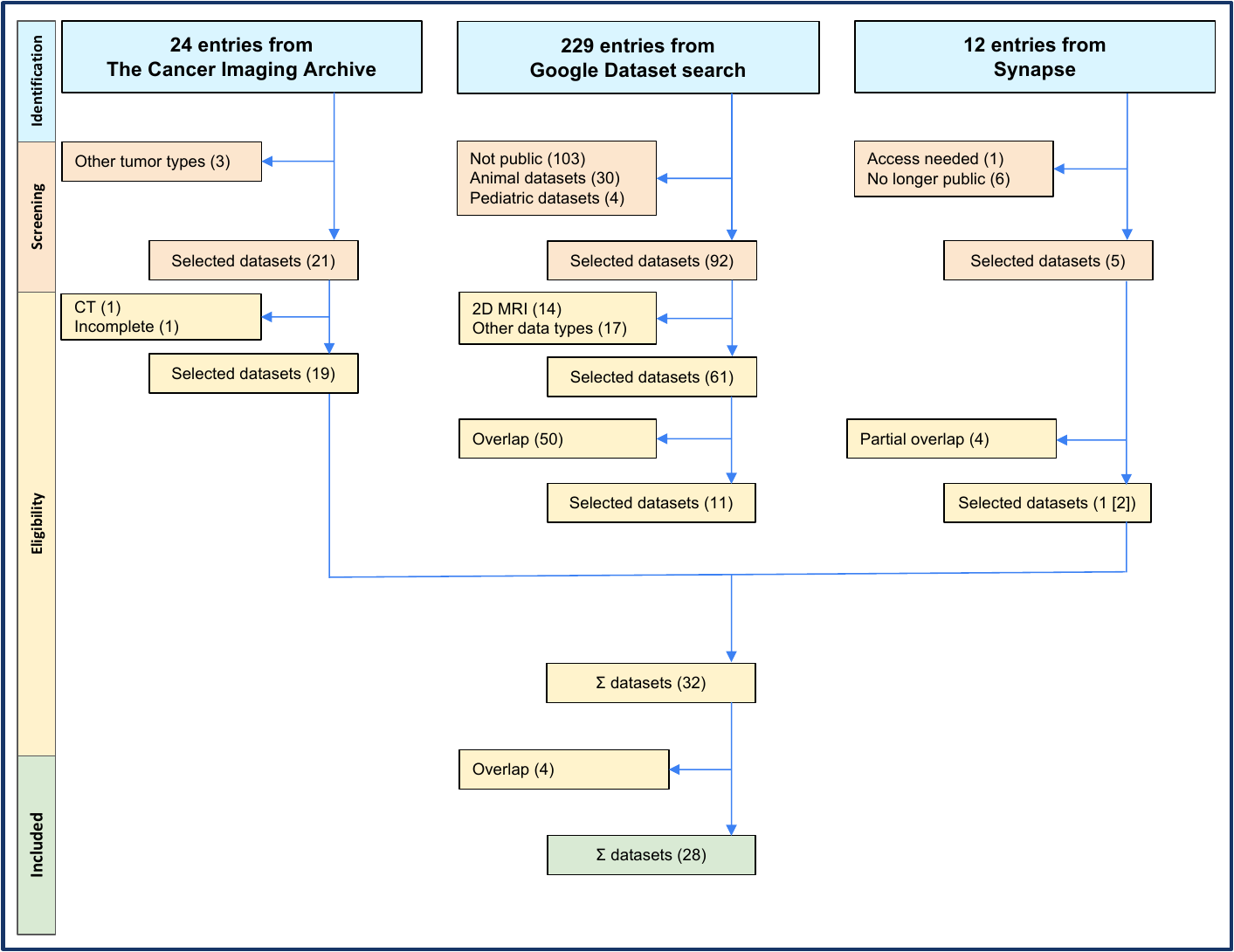}
    \caption{PRISMA workflow diagram for the systematic analysis of public glioma MR Imaging datasets search}
    \label{PRISMA}
\end{figure*}

We followed a PRISMA (Preferred Reporting Items for Systematic Reviews and Meta-Analyses \cite{Moher}) workflow detailed in the flowchart in figure \ref{PRISMA}, which was slightly modified to fit our unique situation where datasets, not studies, are the final object of interest. 

To reveal publicly accessible glioma MRI datasets we searched for datasets using two different search methodologies, a direct and an indirect search. For the direct search a dataset search engine, namely the \textit{Google Dataset search} \footnote{\url{https://datasetsearch.research.google.com}} was queried with the terms ``Glioma'' and ``MRI''. The indirect search was performed on two common dataset archives, \textit{The Cancer Imaging Archive} (TCIA) \footnote{\url{https://www.cancerimagingarchive.net}} and \textit{Synapse} \footnote{\url{https://www.synapse.org/}} using the terms ``Brain'' and ``Glioma'' respectively. 
There were no date restrictions in these searches. During the PRISMA screening phase, datasets from the Google dataset search were excluded if they pertained to animals, were pediatric, consisted of 2D images, lacked MRI images, or were not publicly accessible. Duplicates were also removed. 

To verify that this approach covered relevant datasets, we conducted a validation experiment by looking for dataset references in published articles obtained via \textit{pubmed}, \textit{medarXiv} and \textit{arXiv} using the keywords ``Glioma'', ``MRI'', and ``Dataset''. For PubMed we limited the verification analysis to the first 50 papers returned. This search yielded no additional datasets found, indicating that our search strategy achieved a thorough coverage.

Consequently, 28 datasets remained and were included in the study. The BraTS 2023, which is two subsets \textit{BraTS Adult Glioma} and \textit{BraTS Africa}, is the only one of the BraTS challenge datasets further considered in this review due to BraTS dataset inclusion relations discussed in section \ref{sec:overlaps}.
Note that the field of medical imaging is evolving fast, and thus this review provides a snapshot of the adult glioma MRI dataset state until May 2024.

\section{Results}
\label{sec:result}
4-letter acronyms for each dataset are introduced in Table \ref{acro} to enhance the accessibility and readability of our paper. 
\newcommand{\cmark}{\ding{51}}%
\newcommand{\xmark}{\ding{55}}%

\begin{table}[h]
  \centering
  \caption{The table lists the full name of each dataset along with its corresponding acronym. Four datasets names were not altered as they already contained less than 4 letters (BITE, BTC1, BTC2 and EGD).
The full names of the datasets are according to the TCIA website's collection names. The Test-retest Reliability Data is the name of the file linked to the corresponding paper.}
  \label{acro}
  \begin{small}
  \begin{tabular}{@{}ll@{}}
    \toprule
    \textbf{Dataset} & \textbf{Acronym} \\
    \midrule
    ACRIN-DSC-MR-Brain \cite{boxerman2013early}  & ADMB  \\
    ACRIN-FMISO-Brain \cite{gerstner2016acrin}  & AFMB  \\
    Brain Images of Tumors for Evaluation database \cite{mercier2012online} & BITE  \\
   Brain-Tumor-Progression \cite{schmainda2018data}& BTUP  \\
    BraTS 2023 Adult Glioma \cite{baid2021rsna} & BRAG  \\
     BraTS 2023 Sub Saharan Africa \cite{adewole2023brain} & BRSA  \\
    Brain Tumor Connectomics Data Pre-operative data \cite{aerts2018modeling} & BTC1  \\
    Brain Tumor Connectomics Data Post-operative data \cite{aerts2020modeling} & BTC2  \\
    Burdenko-GBM-Progression & BGBM  \\
    CPTAC-GBM \cite{national2018radiology} & CGBM  \\
    Diffuse Low-grade Glioma Database \cite{parisot2016probabilistic} & DLGG  \\
    Erasmus Glioma Database \cite{van2021erasmus}& EGD  \\
    GLIS-RT \cite{shusharina2021cross} & GLRT  \\
    IvyGAP \cite{puchalski2018anatomic} & IGAP  \\
    IvyGAP-Radiomics \cite{pati2020reproducibility} & IRAD  \\
    LGG-1p19qDeletion \cite{akkus2017predicting} & LGGD  \\
            {LUMIERE \cite{suter2022lumiere}} & LUMI  \\  
    QIN-BRAIN-DSC-MRI \cite{schmainda2018multisite} & QBDM  \\
    QIN GBM Treatment Response \cite{prah2015repeatability} & QGTR  \\
    {REMBRANDT \cite{scarpace2019data}} & REMB  \\
    {RHUH-GBM \cite{cepeda2023r}} & RGBM  \\
    {RIDER NEURO MRI \cite{barboriak2015data}} & RIDN  \\    
    {TCGA-GBM \cite{scarpace2016cancer}} & TGBM \\
    {TCGA-LGG \cite{pedano2016cancer}} & TLGG \\
    {Test-retest Reliability Data \cite{morrison2016reliability}} & TRTR  \\
    {UPENN-GBM \cite{bakas2022university}} & UGBM \\
    {UCSF-PDGM \cite{calabrese2022university}} & UPDG \\
    {ReMIND \cite{juvekar2024remind}} & RMND \\

    \bottomrule
  \end{tabular}
  \end{small}

\end{table}

\subsection{General overview of glioma MRI public datasets}
\begin{figure*}
\includegraphics[width=\textwidth]{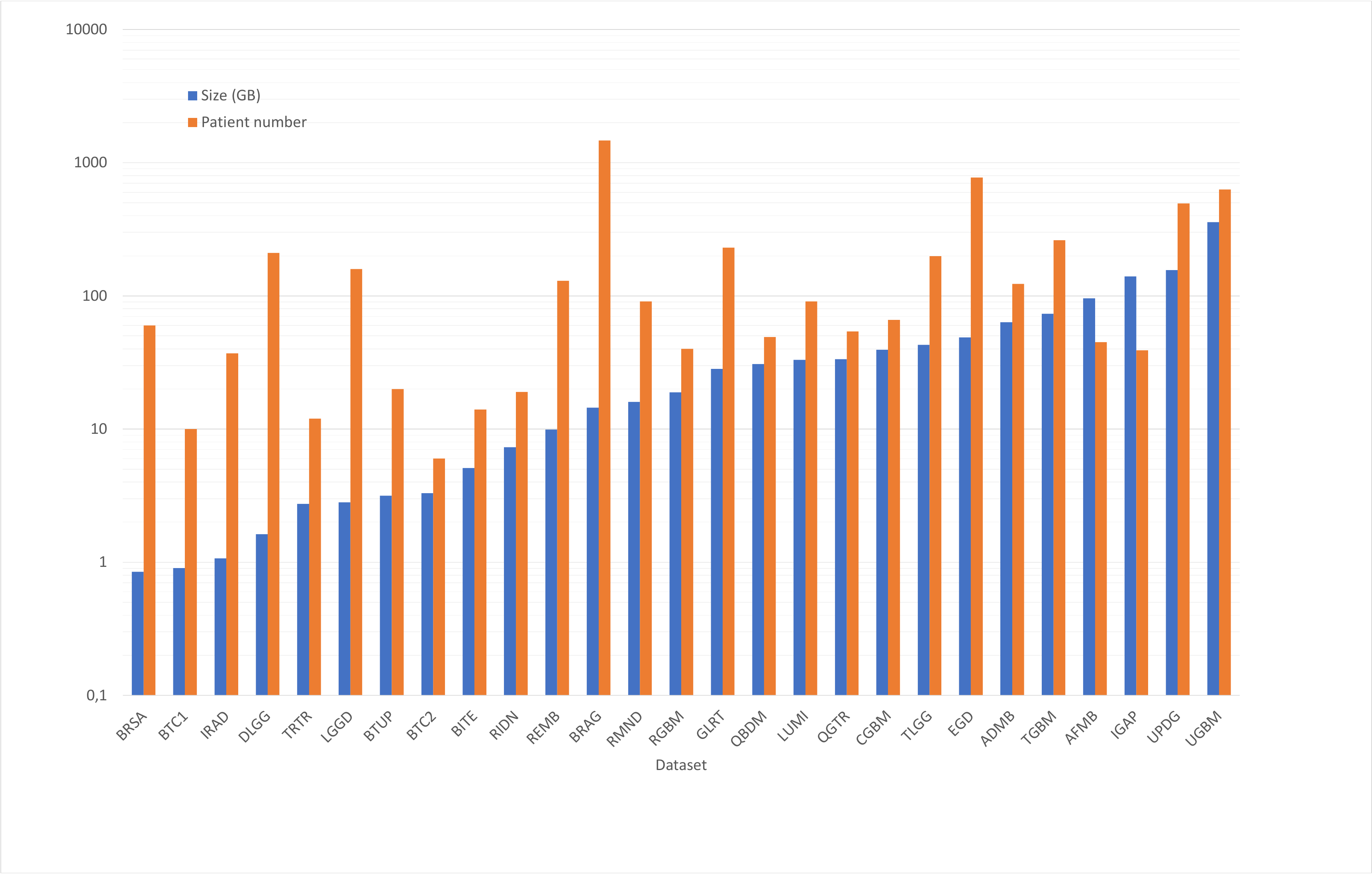}
    \caption{Glioma datasets classified from lowest to highest storage size. Values are shown in logarithmic scale for visualization purposes. BRSA has the smallest size with less than 1 GB (903 MB), while the largest dataset of the study is UGBM with a size of 358 GB.}
    \label{Storage}
\end{figure*}

The greater goal of all these datasets is to help advance the medical cancer research field through medical image analysis. However, they were acquired and released in different contexts and for different primary purposes. Table \ref{overviewtable} introduces the collected datasets, their respective focuses and the journals that published their related works. Figure \ref{Storage} shows the volume in gigabytes and the patient number of the datasets as an overview of their size.

The ADMB and AFMB were integral components of separate \textit{American College of Radiology Imaging Network} (ACRIN) protocols, studying the roles of perfusion MRI, MR spectroscopy, and FMISO PET in the treatment response and survival of GBM patients. QGTR and QBDM are affiliated with the \textit{Quantitative Imaging Network }(QIN) initiative, while the images from TGBM and TLGG are part of more extensive projects of \textit{The Cancer Genome Atlas} (TCGA) focused on understanding genomics. 

Among the 28 selected datasets, 21 were associated with published papers, and 7 datasets lack additional publications but are available directly on the TCIA website. Several datasets, including RMND, LUMI, EGD, BITE, UPDG, RGBM, and UGBM, have dedicated papers explaining their contents, see Table \ref{overviewtable}. \\ \\
The gathered datasets were released for various primary purposes. For example, BTC1, BTC2, and TRTR were involved in comparative studies between glioblastoma patients and healthy controls. Some datasets, such as EGD, DLGG, IRAD, and GLRT, were mainly published to enhance image segmentation methods while others focus on longitudinal problems such as the brain shift and post-surgical tumor segmentation (RMND) .
The purpose of some datasets was to include less conventional MRI modalities, like diffusion and perfusion (e.g., UPDG), while others used these modalities to study the repeatability of perfusion measurements across institutions and patients (QGTR, QBDM, ADMB). Additionally, some datasets were designed with the primary objective of linking imaging data with other types of data, such as genomic, proteomic, and clinical information (LGGD, REMB, CGBM, TGBM, TLGG).  

\begin{table*}[hp]
\caption{Overview of publicly available glioma MRI datasets. The revision of the WHO classification of tumors applied in each dataset is not always available. The AFMB, EGD, LGGD, LUMI, RMND and UPDG are the only ones with clear mention of the WHO classification year. WHO revision years in parenthesis are \textit{estimated} based on the dataset and corresponding paper publication dates. To avoid any wrong affirmation, the WHO year of the BGBM, CGBM, and RIDN was not added.\\
$\star$ Published separately
}
\small

\renewcommand\arraystretch{1.5}
\resizebox{\textwidth}{!}{
\begin{tabular}[t]{lp{9cm}cc}
\toprule
{\textbf{Dataset}} & {\textbf{Focus}} & {\textbf{Journal / Conference}} & {\textbf{WHO revision}}\\
\midrule
{ADMB} & {Role of perfusion MRI and MR spectroscopy in early treatment response in patients receiving bevacizumab} & Neuro-Oncology& (2007) \\
{AFMB } & {FMISO PET and perfusion imaging (Ktrans, CBV) as predictors of survival in GBM}  & Clinical Cancer Research &  2007 \\
BITE & {Development and validation of new image processing algorithms$^{\star}$} & Medical Physics& (2007)  \\

{BTUP}  & {Deep learning for tumor progression prediction}  & Journal of Digital Imaging & - \\

{BraTS}  & {Brain Tumor Segmentation Challenge}  & - & (2007) \\

{BTC1 }  & {Variability of brain activity model parameters between brain tumor patients and healthy controls} &NeuroImage& (2007)\\

{BTC2 } & Changes in model parameters from pre-to post-operative assessment& eNeuro& (2007)  \\

{BGBM}  & {Systematic data collection$^{\star}$} & -&-\\

{CGBM}  & Cancer phenotypes correlation with proteomic, genomic and clinical data  & -&- \\

DLGG& {Tumor segmentation methods and preferential localizations} & PLoS One & (2007) \\

EGD  & {Tumor grading and classification$^{\star}$} &\makecell{Data in Brief} &\makecell{2016}\\

{GLRT} & {Cross-Modality Brain Structures Image Segmentation}  & MICCAI 2020 & (2016)  \\

{IGAP}& {Comprehensive diagnostic characterization of the tumor heterogeneity}  & Science& (2007 or 2016)\\

{IRAD} & {Multi-reader segmentation of GBM tumor}  & Medical Physics Dataset& (2007 or 2016)\\

{LGGD} & {Predicting 1p/19q Deletion in Low-Grade Gliomas}  & Journal of Digital Imaging  & 2007 \\
{LUMI} & Systematic data collection$^{\star}$  & \makecell{Nature: scientific data}& 2016 \\
{QBDM} & {Multisite/multiplatform analyses of DSC-MR imaging datasets}  & \makecell{American Journal \\of Neuroradiology} & (2007 or 2016)\\

{QGTR} & {Repeatability of relative CBV measurements in newly diagnosed glioblastoma}  & \makecell{American Journal \\of Neuroradiology}& (2007)\\

{REMB}  &{Connecting clinical information and genomic data}  &- &(2007) \\

{RGBM}  & {Systematic data collection$^{\star}$} & Data in Brief & (2016 or 2021)\\

{RIDN}  &{Harmonize data collection and analysis for quantitative imaging}  & -& 2000 \\

{RMND}  &{Resource for computational research in brain shift and image analysis}  & Scientific data& 2021 \\

{TGBM}  & {Connecting phenotypes to genotypes using TCGA clinical images}  & -& (2007) \\

{TLGG} & {Connecting phenotypes to genotypes using TCGA clinical images}  & -& (2007) \\

{TRTR} & {Activation map quality divergence between brain tumor patients and healthy controls} & PLoS One& (2007)\\

{UGBM}  & {Systematic data collection$^{\star}$} &Scientific data& (2016) \\

{UPDG} & Preoperative MRI scans with advanced diffusion and perfusion imaging$^{\star}$  & \makecell{Radiology: Artificial \\Intelligence}& 2021 \\

\bottomrule
\end{tabular}
}
\label{overviewtable}
\end{table*}

This comprehensive approach highlights the diverse objectives and applications of the examined datasets in advancing our understanding of glioma imaging and analysis. In the following analysis, we have chosen to focus on the properties of the datasets that are relevant to common medical imaging and, in particular, machine learning approaches.

\subsection{Patient number}
In total the datasets gather 5515 patients where the BRAG, the EGD and the UGBM account for approximately 26.6 $\%$, 14.03 $\%$ and 11.4 $\%$ of the total patient number. The BGBM, TLGG, TGBM, DLGG, GLRT, LGGD, REMB, ADMB and the UPDG cover between 2.2 and 8.9 $\%$ The rest of the datasets cover less than 2 $\%$ of the total patient number. Note that these numbers do not account for potential overlaps between datasets, which are analyzed in the next sections. 

\subsection{Dataset overlaps}
\label{sec:overlaps}
We found that some datasets have (inclusion) relationships, that have to be carefully considered in studies based on this data, for example to avoid model bias by undetected double inclusions of data, leading to data leakage.
Such a relationship exist between the IRAD and the IGAP, where IRAD contains the pre-operative MRIs of the IGAP datasets with additional segmentations and derived radiomics parameters, however the IGAP dataset is longitudinal, while the IRAD is not. We note that the IRAD contains 2 less patients than the IGAP (37 instead of 39). 

The well known BraTS challenge datasets are a special case with respect to complex inclusion relationships that evolved over time. For example \textit{BraTS 2021} contains data that was available in the previous BraTS challenges and other public datasets. \textit{BraTS 2023} further expands on that by extending to 6 different dataset parts, which are used in 9 different challenges \cite{li2023brain}. Here, the ``Adult Glioma'' sub-dataset \textit{is} the BraTS 2021 dataset. The other sub-datasets mostly include non-glioma or paediatric data, except for the \textit{BraTS Sub-Saharan Africa} subset, which is in turn a collection of new glioma imaging data from the african continent.

\begin{figure}
    \centering
    \includegraphics[width=\columnwidth]{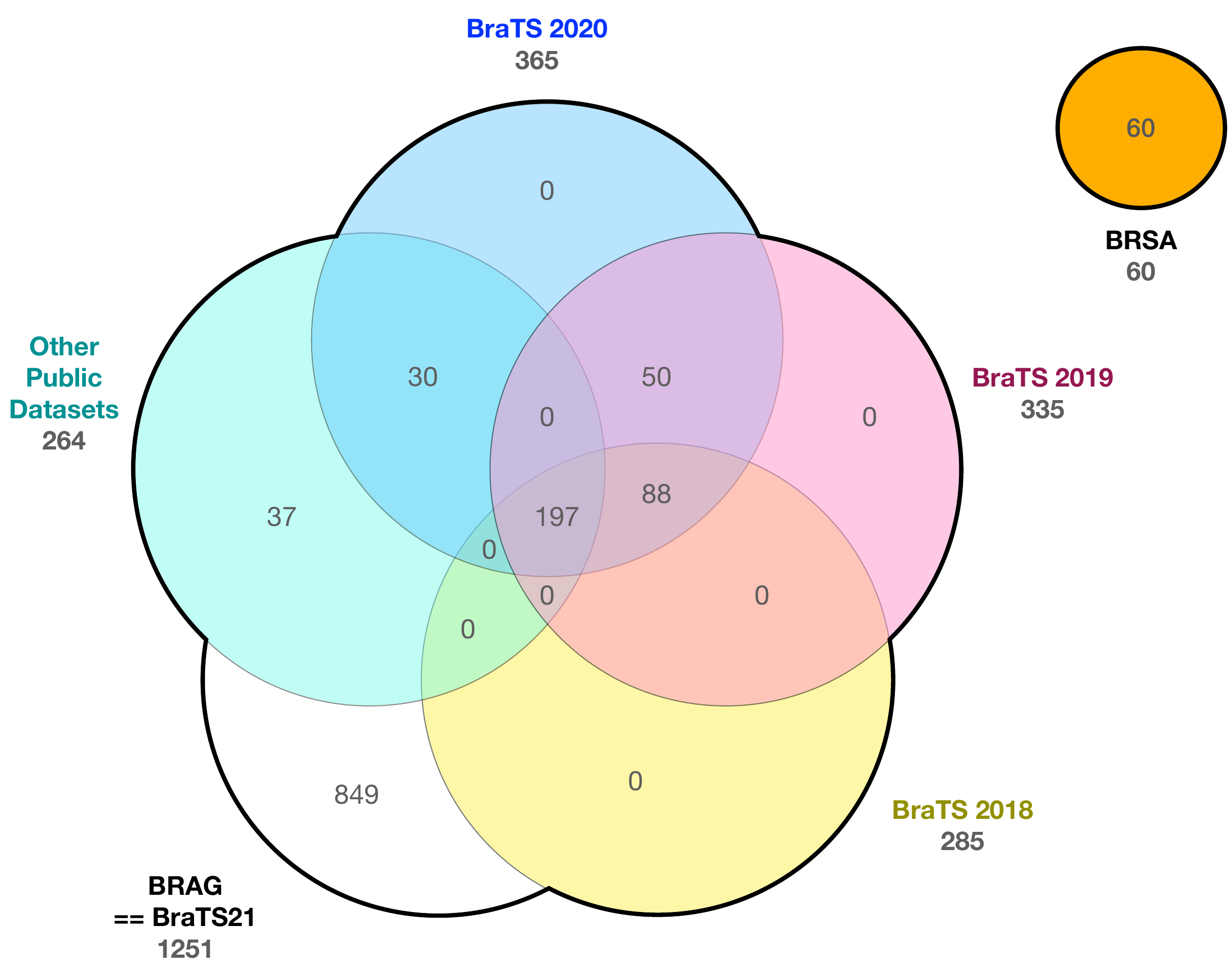} %
    \caption{Adult glioma data inclusion relationships over time in BraTS training datasets based on BraTS 2023, which consists of the BRAG and BRSA datasets. Other public datasets include AFMB, TLGG, TGBM, CGBM and BraTS 2013. BRSA is a new dataset. 
    }
    \label{venn}
\end{figure}

BraTS 2021 added 849 new patients compared to the previous version. The overall inclusion relationships for the versions of BraTS are shown in Figure \ref{venn}.
    \label{brats2021}
The Brats 2021 includes multiple patients from previous datasets: in total, 365 are in BraTS 2020, 335 in BraTS 2019, 285 in BraTS 2018, 285 in BraTS 2017 and 264 from other public datasets. More specifically, 30 are in BraTS 2013, 65 are in TLGG, 102 are in TGBM, 30 in IGAP as can be seen in Figure \ref{brats2021}. 
\\ For consistency purposes, the BraTS 2024 dataset was excluded from this search as the challenge is still ongoing. The exact inclusions along with the validation dataset were not published yet.



\subsection{Dataset release and update dates}
\begin{figure*}[htb]
\centering
\includegraphics[width=\textwidth]{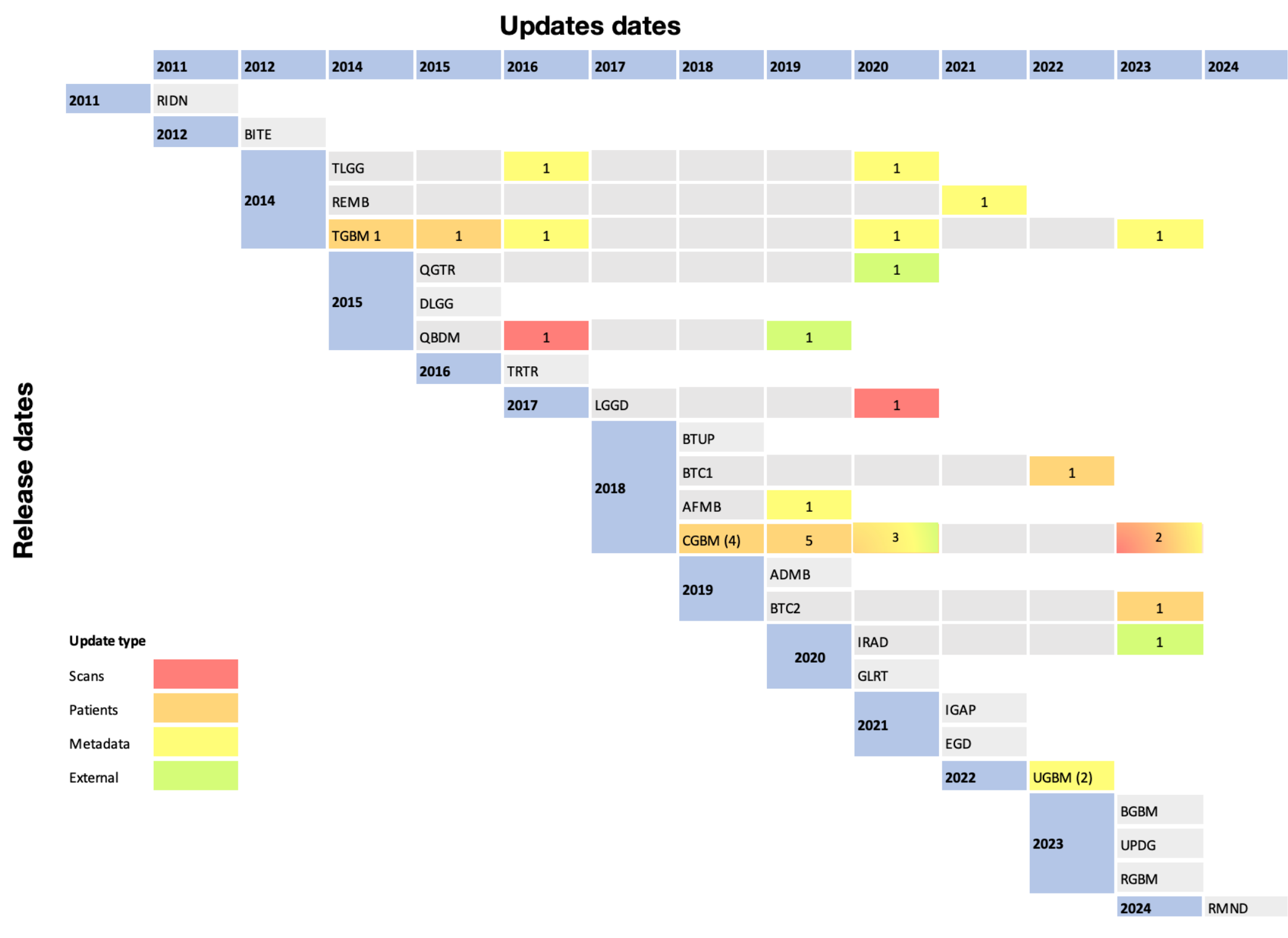}  
    \caption{Classification of datasets based on their release and update dates. Note the absence of new dataset releases in 2013. The table provides the frequency of updates per dataset annually, employing a color-coded system for clarity. A select number of datasets underwent multiple types of updates within a single year; for instance, the CGBM dataset received updates related to patients, metadata, and external sources in 2020, and updates concerning scans and patients in 2023. }
    \label{updates}
\end{figure*}

The datasets in figure \ref{updates} are organized from the most recently released (RGBM and BGBM and UPDG in 2023) to the least recent (RIDN in 2011). Since their initial release dates, $12$ datasets were updated. The CGBM stands out by being updated $14$ times from $2018$ until $2021$. We categorized the updates into four distinct types: scans, patient information, metadata files, and external parameters. These updates included lifting access restrictions, modifying file paths, and altering downloaders. Figure \ref{updates} summarizes the various updated datasets categorized by the type of update. Note that a dataset may appear in multiple sections if it has undergone different types of updates.\\
\\
\textbf{Scans - }The QBDM dataset added six new series in its second update, while the LGGD dataset improved the published segmentations and changed the data format from NIfTI to DICOM. The CGBM dataset received a general data cleanup to remove extraneous scans. In the TGBM dataset, a DICOM tag was repaired in five series for one patient. Finally, 30 DWI MRIs from patients in the AFMB dataset were removed due to inconsistencies in b-value acquisition between GE and Siemens scanners, preventing the reconstruction of ADC maps.\\
\\
\textbf{Patient - }The BTC1 and BTC2 excluded 6 patients, whereas the CGBM (v1 to v12) and the TGBM (v2) added more patients.\\
\\
\textbf{Metadata - }In the AFMB (v2) a new clinical metadata file including the age, treatment and health condition information was added along with some row corrections. One patient tumor type was corrected in CGBM (v12) while more general updates were mentioned for TGBM and TLGG. The only update to the UGBM dataset was related to adding histopathology NDPI slides and updating CSV file for mapping Radiology subject IDs to Histopathology patients. As this change might be related to imaging data, we choose to put it with the metadata update as we only consider MRI data updates in the scans paragraph.\\ 
\\
\textbf{External - } The external update type is the one with less impact as it refers to changes in the dataset format or access methods, rather than modifications of the actual content of the dataset. This type of update does not alter the information contained in the dataset itself. For example, the access embargo was lifted in QGTR, the download link for histopathology slides in the CGBM (v13) was changed and the download location was altered for some files in the IRAD.

\subsection{Dataset WHO classification date}
\begin{figure}
    \centering
    \includegraphics[width=\columnwidth]{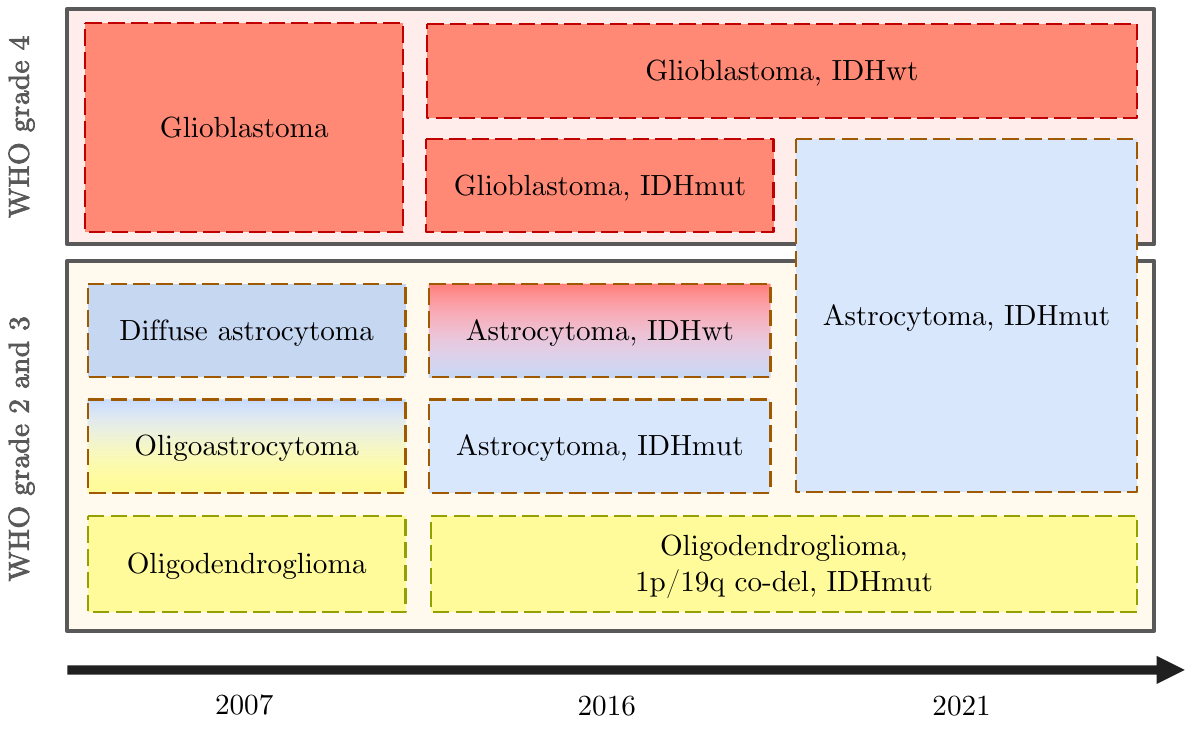}
    \caption{The development of tumor classifications over the last three WHO revisions for the main classes of glioma tumors. Note that astrocytoma IDH wildtype are considered glioblastoma IDH wildtype since the 2021 revision.}
    \label{WHOgrade}
\end{figure}

The 28 datasets have been collected between 2005 and may 2024. Gliomas are classified into different grades per the World Health Organization (WHO). The initial classification was introduced in 1979 with editions in 1993, 2000, 2007 \cite{Louis07} (updated 2016 \cite{Louis16}) and 2021 \cite{louis20212021}. These subsequent updates enable the classification to evolve with main research breakthrough discoveries namely the 1p/19q chromosome codeletion and the Isocitrate DeHydrogenase (IDH) mutation status \cite{reifenberger1994molecular, parsons2008integrated}, which lead to substantial changes in the 4th and 5th editions published in 2007/2016 and 2021.
Figure \ref{WHOgrade} provides an overview of the main changes of glioma type classification since the introduction of molecular diagnostics \cite{barthel2018evolving}. 

The WHO grading system includes grade 2, 3 and 4\cite{van2018grade}. In the WHO 2021 revision, grade 2 and 3 include two subtypes: oligodendrogliomas (IDH-mutant and 1p/19q chromosome codeleted) and astrocytomas (IDH-mutant with no 1p/19q codeletion). Grade 4 tumors are categorized into astrocytoma (IDH-mutant) and glioblastoma (IDH-wildtype) as the glioblastoma IDH-mutant type does not exist anymore (Fig.\ref{WHOgrade}). \\ 
Among the datasets collected, only AFMB (WHO 2007), LGGD (WHO 2007), EGD, LUMI (WHO 2016), the RMND and UPDG (WHO 2021) explicitly state the specific WHO edition version employed for tumor classification. For the remaining datasets, the WHO edition was inferred based on the publication date and associated articles, where available. 
More specifically, to avoid any wrong affirmation, the WHO year of the BGBM, CGBM, and RIDN was not added in this study (Table \ref{overviewtable}). 

Finally, the majority of the datasets are believed to follow the WHO 2007 with respect to tumor classification. However this information cannot be confirmed without contacting the authors of the corresponding datasets. 

\subsection{Longitudinal studies}
\begin{figure}
    \centering
    \includegraphics[width=\columnwidth]{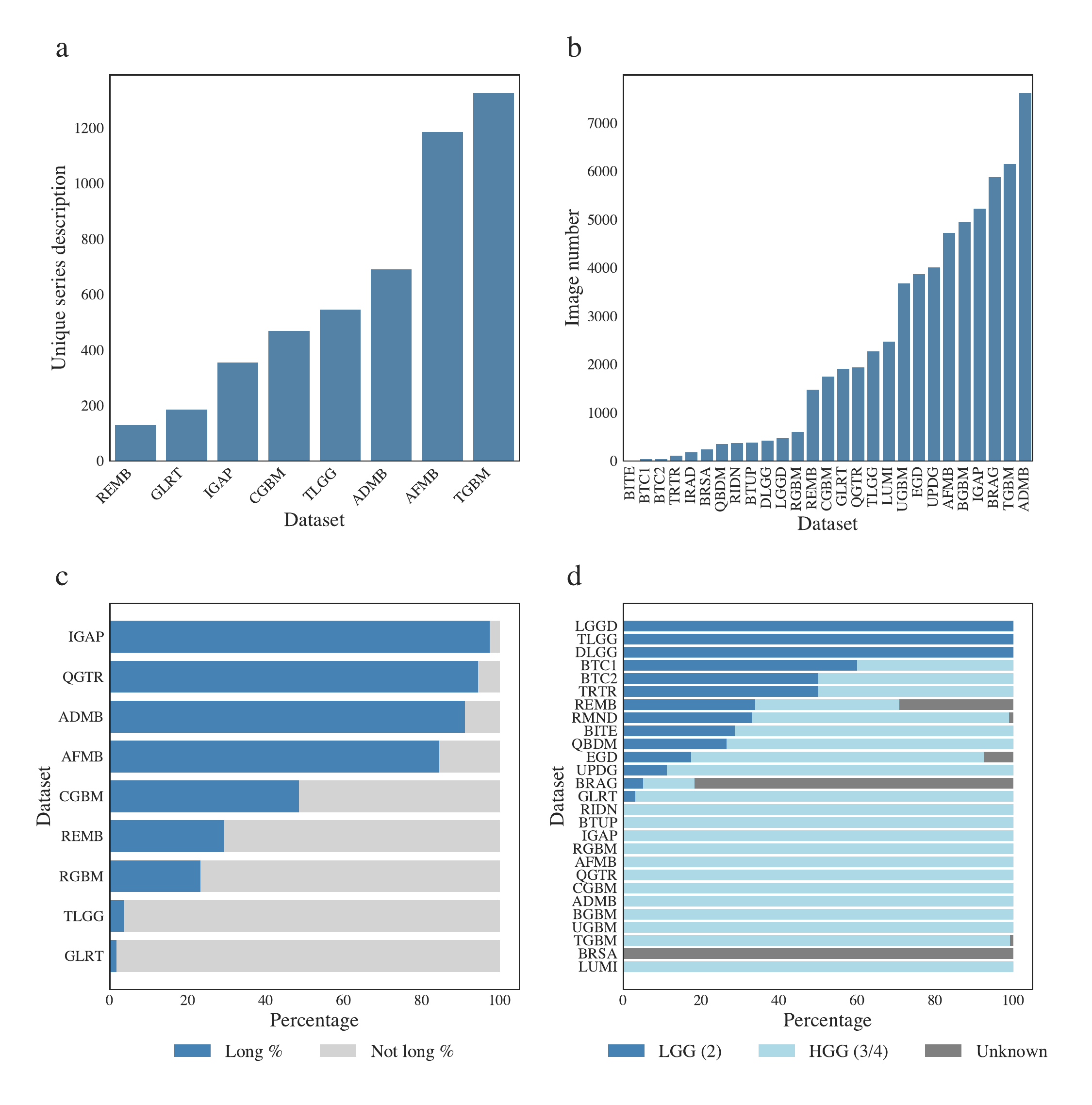}
    \caption{\textbf{(a)} Number of distinct MRI series descriptions of the 8 remaining datasets  \textbf{(b)} Total image number per dataset \textbf{(c)} Percentage of longitudinal studies across mixed datasets. Long $\%$ stands for percentage of longitudinal datasets \textbf{(d)} Percentage of tumor grades across datasets}
    \label{statdata}
\end{figure}

Out of the 28 datasets, 13 are cross-sectional, 6 are fully longitudinal (LUMI, BGBM, RGBM, BTUP, RIDN, RMND), and the remaining 9, mixed datasets containing both cross-sectional and longitudinal studies, are shown in Figure \ref{statdata}c. In the mixed datasets the IGAP is the dataset with the maximum percentage of longitudinal data (38 of 39 patients), while the GLRT represents the dataset with the least percentage of longitudinal data (4 of 226 patients). 

Within the longitudinal and mixed datasets, three datasets have the same number of sessions for all longitudinal patients (2 for the BTUP dataset and the RIDN and 3 for the BGBM dataset). 
\\ 
The IGAP contains the maximum number of timepoints for one patient (27 timepoints) followed by the AFMB and the TGBM with 25 and 23 timepoints.
\\
\subsection{Tumor grades}
In the rest of the paper, grade 2 gliomas are considered low grade gliomas (LGG) while grade 3 and 4 gliomas are classified as higher grade gliomas (HGG)
The availability of tumor grade information varied across different datasets. In total we have 58.5\% of HGG, 17\% of LGG, and 24.3\% are labelled as "unknown" including patients from datasets with incomplete or missing tumor grade data. 
\\
The exact tumor grade for BRAG and BRSA is not publicly available. Therefore all 
patients in BRSA were categorized as unknown grade. In contrast, for BRAG, we deduced the grade from a mapping file linking the patients from BRAG to their original source datasets. The naming of these original source datasets often included the keywords “LGG” or “HGG” indicating the tumor grades allowing us to infer the tumor grades. Through this method, we were able to identify that at least 193 patients with HGG and 75 patients with LGG in the BRAG dataset.\\

Additionally, it is worth noting that the patient number and associated tumor grades and types of TRTR was taken from the associated published paper as it was difficult to deduce it from the dataset alone. The tumor grades and types of the QBDM was also extracted from the associated paper. In the 28 datasets, we counted 944 LGG (grade 2) and 3233 HGG (grade 3 and 4) and 1344 unknown grades as shown in Figure  \ref{statdata}d . 

\subsection{Tumor types}
According to the WHO classification, glioma tumor types are identified according to the IDH mutation and the 1p19q codeletion information. \\Regarding the former, the IDH status is easily available for all patients of the UPDG and the RGBM datasets and is partially available for the EGD (467/774 patients), the UGBM (515/630 patients), the BGBM (114/180 patients) and the LUMI (58/91 patients) datasets. \\
The 1p19q codeletion on the other hand, is only clearly available for all patients of the LGGD datasets, and partially for the EGD (259/774 patients), and the UPDG (405/630 patients) datasets.
\\
Both histopathological and molecular classification criteria are less clear in other datasets which either do not provide them or require a search on the website of the dataset provider. This concerns datasets such as the TLGG, TGBM, or IGAP/IRAD.
\\
Furthermore, the specific tumor type may be guessed from the IDH mutation and 1p19q codeletion information if available. For example, an IDH mutant tumor with no 1p19q codeletion may be denoted as astrocytoma. As such, one can conclude on the WHO 2021 corresponding classification. This process is possible in few datasets (EGD, UGBM, UPDG, BGBM and RGBM). On the contrary, when the dataset contains the tumor type along with the exact WHO revision, the IDH mutation and 1p19q codeletion can be deduced (RMND).
%
\subsection{Magnetic Resonance Images in the datasets} 
In Table \ref{MRInumb} we focus on MRI modalities and segmentations in 20 datasets for which the information was available in the description.
Note that the numbers of images and of subjects strictly include glioma and MRI. Thus, we consider a subset of BITE (removing ultrasound imaging), BTC1 and BTC2 (removing all non-glioma MRI), RMND (removing ultrasound imaging and all non-glioma MRI) and TRTR (removing healthy control patients).
Observe that $10$ datasets (BRAG, BRSA, BGBM, BTUP, EGD, IRAD, RGBM, UGBM, UPDG) provide all four conventional MRI modalities (T1, T2, T1 Contrast Enhanced (T1CE) and FLAIR). Diffusion and diffusion derived (ADC) information is available in 7 Datasets while 6 datasets provide perfusion information (ASL, DSC, DCE) or corresponding derived information such as the rCBV. 3 Datasets provide fMRI images while SWI and HARDI MRIs are available in UPDG. LGGD provides T2CE MRIs for all patients. \\
The tumor segmentation information is included in 14 datasets: in BITE and DLGG, the masks are in minc and xml format respectively. Both GLRT and BGBM provide the Gross Tumor Volume (GTV) and Contour Tumor Volume (CTV) that are more commonly used in the clinics. The RMND dataset comprises presurgical tumor segmentation, intraoperative residual tumor and automatic segmentations of cerebrum and ventricles for some cases.
\\ \\MRI information in the 8 remaining datasets was not explicitly documented in the general descriptions. To address this gap, we turned to the metadata files associated with these datasets. Unfortunately, due to the high number of unique MRI series description and images (Fig. \ref{statdata}a and \ref{statdata}b), finding the corresponding modalities and segmentations for each of them was out of the scope of this study.  

%
\section{Discussion}
\label{sec:discu}
In this section, we discuss four main points: 1) the practical challenges imposed by the WHO classification updates in the datasets, 2) different problem specific quality criteria applied across datasets, 3) implications of specific research questions on dataset choice and finally, 4) limitations of this review. 


\subsection{Importance and practical implications of the WHO classification} 
The WHO's updates to the central nervous system tumor classification edition, driven by increased understanding of molecular factors, has significant and immediate implications for radiologists and neuropathologists \cite{johnson20172016}, \cite{martinez2018practical}. 
Nevertheless, the practical implementation of these changes in the medical field is not immediate, as it requires the adaptation of jargon and classification systems that have been used daily for more than 25 years since the first edition was published in 1979 \cite{zulch1979histological,nakazato2008revised}. 

The nature of this process may appear insignificant, yet its implications for patient care, clinical trials and training of artificial intelligence models is undeniable. Specifically, the transition delay could significantly impact public glioma datasets and all downstream development done based on them. 
For example, very popular and large datasets appear to still adhere to the 2007 edition of the WHO classification. Consequently, machine learning models trained on them will produce diagnoses aligned with the 2007 standard. In particular, such models - like the ones trained on BraTS - would likely misclassify grade 4 astrocytomas as glioblastomas (cf. Figure \ref{WHOgrade}). 

Updating the classification labels in existing datasets is often infeasible due to the absence of molecular and/or histopathological tests. For example, LGGD adheres to the 2007 revision. However, even though the 1p19q deletion status is included and would suggest oligodendroglioma, the unknown IDH mutation status prevents us from confidently providing an updated WHO classiﬁcation. The EGD dataset is an additional example, with 291 patients lacking both the IDH mutation and 1p19q codeletion information. Another problem regarding WHO classification updates lies within longitudinal datasets. For patients with recurrent tumors and long follow-up, the WHO classification might have changed during the course of disease. None of the longitudinal datasets was updated accordingly which might affect longitudinal studies of treatment response. 
Also, 61 $\%$ of the tumors in TLGG have been classified as "oligoastrocytoma" which are tumors considered as a mixture of cells that originated from oligodendrocytes and astrocytes. 
This term was removed in WHO revision in 2016, and additional information would be required to re-classify these tumor types (1p19q codeletion and IDH mutation). 

\begin{table*}[htbp]
\centering
\renewcommand\arraystretch{1.5}
  \caption{Adult glioma MRI modalities and segmentations available per dataset. (P) partially available. The datasets with unclear MRI descriptions were excluded from this table. Note that some data sets contain additional non-glioma data not counted here.
  }
 \resizebox*{\textwidth}{!}{%
  \begin{tabular}{@{}l r r c c c c c c c c c c c c c c@{}}
    \toprule
    \textbf{Dataset} & 
    \makecell{\textbf{No.}\\\textbf{images}} & 
    \makecell{\textbf{No.}\\\textbf{subjects}} & 
    \multicolumn{7}{c}{\textbf{MRI modality}} &  \phantom{a} & 
    \multicolumn{2}{c}{\textbf{Segmentation}} & 
    \phantom{a} &
    \multicolumn{3}{c}{\textbf{Preprocessing}} \\
    \cmidrule{4-10} \cmidrule{12-13} \cmidrule{15-17}
             &          &          & T1 & T1CE  &T2&FLAIR&Diffusion&Perfusion&Additional    & & Tumor& Additional & & Format & Skull strip. & registered  \\
    \midrule
    BGBM    & 4956  & 180  & \textcolor{OliveGreen}\cmark     & \textcolor{OliveGreen}\cmark          & \textcolor{OliveGreen}\cmark  & \textcolor{OliveGreen}\cmark &\textcolor{red}\xmark &\textcolor{red}\xmark &\textcolor{red}\xmark 
    & &\makecell{\textcolor{OliveGreen}\cmark (\footnotesize{GTV-CTV}\\\footnotesize{-PTV})}
    &\textcolor{OliveGreen}\cmark 
    & & \footnotesize{DICOM} & \footnotesize{No} & \footnotesize{No}
    \\
    \hline
        BITE    & 14  & 14  & \textcolor{red}\xmark     & \textcolor{OliveGreen}\cmark   & \textcolor{red}\xmark & \textcolor{red}\xmark  &\textcolor{red}\xmark &\textcolor{red}\xmark &\textcolor{red}\xmark 
    & &\textcolor{red}\xmark 
    & \textcolor{red}\xmark 
    & & \footnotesize{MINC} &\footnotesize{No} &     \footnotesize{Yes, MNI}\\
        \hline
        BRAG    &  5880  & 1470 & \textcolor{OliveGreen}\cmark & \textcolor{OliveGreen}\cmark   & \textcolor{OliveGreen}\cmark & \textcolor{OliveGreen}\cmark & \textcolor{red}\xmark & \textcolor{red}\xmark &\textcolor{red}\xmark 
        & &\makecell{\textcolor{OliveGreen}\cmark \footnotesize{3-label}} &\textcolor{red}\xmark 
& &\footnotesize{NIfTI} &\footnotesize{Yes} & \footnotesize{Yes, SRI}\\
        \hline
        BRSA  &  240 &  60 & \textcolor{OliveGreen}\cmark & \textcolor{OliveGreen}\cmark   & \textcolor{OliveGreen}\cmark & \textcolor{OliveGreen}\cmark&\textcolor{red}\xmark&\textcolor{red}\xmark &\textcolor{red}\xmark
        & &\makecell{\textcolor{OliveGreen}\cmark \footnotesize{3-label}} &\textcolor{red}\xmark
        & &\footnotesize{NIfTI} &\footnotesize{Yes} & \footnotesize{Yes, SRI}\\
        \hline
        BTC1    & 40  & 10  & \textcolor{OliveGreen}\cmark & \textcolor{red}\xmark   & \textcolor{red}\xmark & \textcolor{red}\xmark  &\makecell{\textcolor{OliveGreen} \cmark \footnotesize{(DWI)}}&\textcolor{red}\xmark &\makecell{\textcolor{OliveGreen}\cmark \footnotesize{(BOLD)}}
        & &\textcolor{red}\xmark &\textcolor{red}\xmark  & &\footnotesize{NIfTI} &\footnotesize{No} & \footnotesize{No}\\
               \hline
        BTC2    & 40  & 10  & \textcolor{OliveGreen}\cmark & \textcolor{red}\xmark   & \textcolor{red}\xmark & \textcolor{red}\xmark  &\makecell{\textcolor{OliveGreen} \cmark \footnotesize{(DWI)}}&\textcolor{red}\xmark & \makecell{\textcolor{OliveGreen}\cmark \footnotesize{(BOLD)}}
        & &\textcolor{red}\xmark &\textcolor{red}\xmark & &\footnotesize{NIfTI} &\footnotesize{No} & \footnotesize{No}\\
                        \hline
        BTUP    & 383  & 20  & \textcolor{OliveGreen}\cmark & \textcolor{OliveGreen}\cmark  & \textcolor{OliveGreen}\cmark & \textcolor{OliveGreen}\cmark  & \makecell{\textcolor{OliveGreen}\cmark \footnotesize{(ADC)}}&\makecell{\textcolor{OliveGreen}\cmark \footnotesize{(nCBF-crCBV}\\ \footnotesize{-srCBV-DSC)}}&\textcolor{red}\xmark 
        & &\textcolor{OliveGreen}\cmark  &\textcolor{red}\xmark  
        & &\footnotesize{DICOM} &\footnotesize{No} & \footnotesize{Yes, intra-subject}\\
        \hline
        DLGG    & 420  & 210  & \textcolor{red}\xmark &  \textcolor{red}\xmark  & \textcolor{red}\xmark & \textcolor{OliveGreen}\cmark  &\textcolor{red}\xmark&\textcolor{red}\xmark&\textcolor{red}\xmark & &\makecell{\textcolor{OliveGreen}\cmark \footnotesize{ (.xml)}}&\textcolor{red}\xmark 
    & &\footnotesize{NIfTI} &\footnotesize{No} & \footnotesize{No}\\
    \hline
        EGD    & 3870  & 774  & \textcolor{OliveGreen}\cmark &  \textcolor{OliveGreen}\cmark & \textcolor{OliveGreen}\cmark & \textcolor{OliveGreen}\cmark  &\textcolor{red}\xmark&\textcolor{red}\xmark&\textcolor{red}\xmark 
        & &\textcolor{OliveGreen}\cmark &\textcolor{red}\xmark 
    & &\footnotesize{NIfTI} &\footnotesize{No} & \footnotesize{Yes, MNI}\\
    \hline
    IRAD    & 185  & 37  & \textcolor{OliveGreen}\cmark &  \textcolor{OliveGreen}\cmark & \textcolor{OliveGreen}\cmark & \textcolor{OliveGreen}\cmark  &\textcolor{red}\xmark&\textcolor{red}\xmark&\textcolor{red}\xmark & &\textcolor{OliveGreen}\cmark &\textcolor{red}\xmark 
    & &\footnotesize{NIfTI} &\footnotesize{Yes} & \footnotesize{Yes, SRI and MNI}\\
    \hline
    LGGD    & 478  & 159  & \textcolor{red}\xmark &  \textcolor{OliveGreen}\cmark & \textcolor{red}\xmark & \textcolor{red}\xmark  &\textcolor{red}\xmark&\textcolor{red}\xmark&\makecell{\textcolor{OliveGreen}\cmark\\(T2CE)}  & &\textcolor{OliveGreen}\cmark &\textcolor{red}\xmark 
    & & \footnotesize{DICOM} &\footnotesize{No} & \footnotesize{No}\\
    \hline
    QBDM    & 349  & 49  & \textcolor{red}\xmark &  \textcolor{OliveGreen}\cmark & \textcolor{red}\xmark & \textcolor{red}\xmark  &\textcolor{red}\xmark&\makecell{\textcolor{OliveGreen}\cmark\\(DSC)}&\textcolor{red}\xmark  & &\textcolor{OliveGreen}\cmark &\textcolor{OliveGreen}\cmark 
& &\footnotesize{DICOM} &\footnotesize{No} & \footnotesize{Yes, intra-subject}\\
    \hline
QGTR    & 1942  & 54  & \textcolor{OliveGreen}\cmark &  \textcolor{OliveGreen}\cmark & \textcolor{OliveGreen}\cmark& \textcolor{OliveGreen}\cmark&\textcolor{OliveGreen}\cmark&\makecell{\textcolor{OliveGreen}\cmark\\(DCE-DSC)}&\makecell{\textcolor{OliveGreen}\cmark\\(MEMRAGE)}  & &\textcolor{red}\xmark &\textcolor{red}\xmark 
& &\footnotesize{DICOM} &\footnotesize{No} & \footnotesize{Yes, intra-subject}\\
    \hline
RGBM    & 600  & 40  & \textcolor{OliveGreen}\cmark &  \textcolor{OliveGreen}\cmark & \textcolor{OliveGreen}\cmark& \textcolor{OliveGreen}\cmark&\textcolor{red}\xmark &\textcolor{red}\xmark &\textcolor{red}\xmark & &\textcolor{OliveGreen}\cmark &\textcolor{red}\xmark 
& &\footnotesize{NIfTI} &\footnotesize{Yes} & \footnotesize{Yes, SRI}\\
    \hline
RIDN    & 368  & 19  & \textcolor{red}\xmark &  \textcolor{OliveGreen}\cmark & \textcolor{red}\xmark& \makecell{\textcolor{OliveGreen}\cmark\\(P)}&\makecell{\textcolor{OliveGreen}\cmark\\(DTI)} &\makecell{\textcolor{OliveGreen}\cmark\\(DCE)} &\textcolor{red}\xmark 
& &\textcolor{red}\xmark &\textcolor{red}\xmark 
& &\footnotesize{DICOM} &\footnotesize{No} & \footnotesize{No}\\
    \hline
TRTR    & 108  & 12  & \textcolor{red}\xmark &  \textcolor{red}\xmark & \textcolor{red}\xmark& \textcolor{red}\xmark&\textcolor{red}\xmark &\textcolor{red}\xmark&\makecell{\textcolor{OliveGreen}\cmark\\(fMRI)}& &\textcolor{red}\xmark &\textcolor{red}\xmark &  & \footnotesize{NIFTI} & \footnotesize{No}& \footnotesize{No}\\ 
    \hline
UGBM    & 3680  & 630  & \textcolor{OliveGreen}\cmark  &  \textcolor{OliveGreen}\cmark  & \textcolor{OliveGreen}\cmark & \textcolor{OliveGreen}\cmark &\makecell{\textcolor{OliveGreen}\cmark \\(P)} &\makecell{\textcolor{OliveGreen}\cmark \\(P)(DSC)}&\textcolor{red}\xmark& &\makecell{\textcolor{OliveGreen}\cmark \\(P)} &\textcolor{red}\xmark 
& & \footnotesize{NIfTI} &\footnotesize{Yes} & \footnotesize{Yes, unknown atlas}\\ 
    \hline
UPDG    & 4008  & 495  & \textcolor{OliveGreen}\cmark  &  \textcolor{OliveGreen}\cmark  & \textcolor{OliveGreen}\cmark & \textcolor{OliveGreen}\cmark &\textcolor{OliveGreen}\cmark &\makecell{\textcolor{OliveGreen}\cmark \\(ASL)}&\makecell{\textcolor{OliveGreen}\cmark \\(SWI\\-HARDI)}& &\textcolor{OliveGreen}\cmark &\textcolor{red}\xmark 
& & \footnotesize{NIfTI} &\footnotesize{Yes} & \footnotesize{Yes, intra-subject}\\
\hline
LUMI    & 2478  & 91  & \textcolor{OliveGreen}\cmark  &  \textcolor{OliveGreen}\cmark  & \textcolor{OliveGreen}\cmark & \textcolor{OliveGreen}\cmark &\textcolor{red}\xmark &\textcolor{red}\xmark& \textcolor{red}\xmark&&\makecell{\textcolor{OliveGreen}\cmark \footnotesize{3-label}} &\textcolor{red}\xmark 
& & \footnotesize{NIfTI} & \footnotesize{Yes} & \footnotesize{Yes, unknown altas}\\ 
\hline
RMND    & 841  & 91  & (P)  &  (P) & (P) & (P) &\textcolor{red}\xmark &\textcolor{red}\xmark& \textcolor{red}\xmark&&\makecell{\textcolor{OliveGreen}\cmark } &\textcolor{OliveGreen}\cmark 
& & \footnotesize{DICOM} & \footnotesize{No} & \footnotesize{Yes, intra-subject}\\ 
    \bottomrule
  \end{tabular} 
  }
\label{MRInumb}
\end{table*}


\subsection{Problem specific quality criteria}
The most appropriate dataset for a specific problem changes depending on the intended use. 
For example, if the study focuses on a specific tumor type, appropriate datasets are limited to sets that include detailed type classification, including IDH mutations status.
For this case the EGD, UGBM, UPDG, BGBM and RGBM datasets would be the preferred choice. 
\\ 
If the intended study focuses on specific differences between low grade and high grade tumors, the imbalance of datasets needs to be considered. As introduced previously, more than $50 \%$ patients across all the datasets have a high grade tumor. 
\\
For observing tumor evolution or treatment effect on tumor volume over time, datasets are restricted to the ones that provide longitudinal MRIs. For that purpose, Figure \ref{statdata}c illustrates the percentage of patients with longitudinal MRIs within all longitudinal datasets. The most adequate choice in that case might be the IGAP dataset.
\\
Besides the WHO classification date, the study date itself might be crucial as well. Overall MRI quality is expected to be better in 2023 than in 1995. This needs to be taken into account when training models intended to have clinical impact. 
\\
Potential population shifts are also to be considered. For example, different healthcare systems across the globe might lead to different stages where MR imaging takes place, different scanner hardware generations used, and different therapy regimes in turn influencing the imaging phenotype. A first dataset considering a specific geographical inclusion criteria is BRSA.
\\
Datasets that include tumor masks along with the intended MRI modality facilitate the training of models for tumor segmentation.

\subsection{Research questions examples and available public datasets}
In this section we discuss the relevance of dataset use in the scope of three different research questions: tumor growth, MRI normalization, and tumor segmentation.

\paragraph{Tumor growth.} One of the central questions in cancer research is understanding the causes of tumor evolution over time. This is necessary for predicting tumor growth, customizing treatments, and potentially preventing the tumor from reaching a critical stage. These research questions require longitudinal patient data. 
In our study, within the mixed datasets, only 4 datasets are primarily longitudinal (IGAP, QGTR, ADMB and AFMB) and all of them include already grade 4 glioblastoma with treatment and resection surgeries. These would be appropriate to answer research questions related to multisite analysis of treatment response. Nonetheless, post-operative MRIs might be present in these datasets affecting tumor growth and heterogeneity prediction. In addition, brain shift makes it very difficult to identify spatial differences in tumor evolution. According to surgeons' opinion, 6 weeks approximately are needed for the brain shift's effect to be negligible \cite{gerard2017brain}. However, the RMND dataset might help tackle this issue as it contains preoperative MRIs along with registered intraoperative MRIs and, in some cases, the residual tumor segmentations.

\paragraph{Modeling of malignant transformation.} Tumor location and size are not the only factors that evolve through time. Tumor heterogeneity and aggressivity also change. More specifically, low grade gliomas with IDH mutation, almost always recur as a higher more aggressive grade through malignant transformation (MT) \cite{nakasu2022malignant,bogdanska2017mathematical}. This process occurs gradually, through changes in the tumor micro-environment \cite{lima2016incidental,bready2019molecular}. However, the reasons leading to MT are not yet fully understood. This is why, using available patient data, mathematical models attempt to describe tumor growth and heterogeneity until the malignant transformation is diagnosed \cite{jalbert2016magnetic}. In this case, longitudinal datasets alone are not sufficient. Additional criteria are needed: IDH mutant grade 2 gliomas with at least three timepoints serving to initialize, calibrate and evaluate a patient-specific mathematical model as described in \cite{hormuth2017mechanically, Urcun2023} and \cite{jarrett2018incorporating} on brain and breast tumors. Furthermore, MT must have been histopathologically confirmed at the third time point. 
In the studied longitudinal datasets, only 1 patient with a grade 2 to 3 astrocytoma was found (TCGA-LGG) that meets this criteria. However, only two imaging sessions were identified, making the evaluation step of mathematical modeling not possible for this patient.
Conventional MRIs (T1, T2, FLAIR, T1CE) are sufficient for tumor segmentation. However, more MRI modalities are needed to inform about the tumor micro-environment during MT, such as the Apparent Diffusion Coefficient (ADC) map describing diffusion of water molecules \cite{ignjatovic2015apparent,wong2021prediction} and the (relative Cerebral Blood Volume) rCBV for tumor vascularization \cite{danchaivijitr2008low, upadhyay2011low,bobek2014anaplastic}. To address this gap, a solution would be to predict the missing MRI modalities, leading to new research questions \cite{islam2021glioblastoma,azad2022medical,khalid2023multimodal}. \\

\textit{Tumor segmentation.} The BraTS datasets are the most prominent ones specifically published for segmentation and were released in a challenge context. 
As such, the MRIs are highly preprocessed: Skull stripped, registered to the same template, in the NIfTI format and resampled to 1 $mmˆ3$. From Table \ref{MRInumb}, 8 datasets may potentially be added to BraTS to train a model for tumor segmentation (BGBM, BTUP, IRAD, LUMI, RGBM, RMND, UGBM and UPDG). 
Segmentation models trained on such datasets may be used for tumor segmentations in datasets that include the 4 classical MRIs such as QGTR but no segmentation maps.  
\subsection{Limitation of the review}
We recognize several limitations of our review that might be addressed in further works. 
Firstly, due to the WHO classification changes, it is challenging to apply the current tumor type specification without information about IDH and 1p19q status in existing datasets, thus we can not provide a complete and detailed overview on the covered tumor types according to the recent 2021 specification.

Secondly, 8 datasets out of the 28 were excluded from table \ref{MRInumb} due to unclear MRI descriptions, which might be partially recoverable by the original dataset authors. 

Lastly, the field is moving fast and new datasets are published regularly. Therefore, this review can only be considered as a snapshot of adult glioma datasets until May 2024. 

\section{Conclusion}
In the context of the data scarcity faced by computational researchers in medical imaging, publicly available datasets represent a valuable asset. In this article, we provided a comprehensive overview of MRI public adult glioma datasets and highlighted their potentials and challenges. The resources would serve as a foundation for researchers studying adult glioma tumors. 

Across the 28 gathered datasets, only the UPDG and the RMND datasets follow the current WHO classification criteria. For the UPDG dataset, all classical MRIs are available along with diffusion and perfusion MRI modalities. A tumor mask is also provided in NIfTI format and the main preprocessing steps are performed (skull stripping and co-registration). Additionally, the corresponding histopathological and molecular characteristics are available along with treatment relevant information such as MGMT methylation status. Regarding RMND, the  classical pre-operative MRIs are available for almost all patients in DICOM format. Intra-operative MRIs are also available. The tumor type along with its corresponding WHO classification are informed for all patients except one.

In Section \ref{sec:discu}, we discussed the possible usage of the different datasets for specific research purposes. A careful selection is crucial, and researchers must define their study objectives precisely. The IGAP, QGTR, ADMB and AFMB datasets are relevant for studying glioblastoma response to treatment, while BGBM, BRAG, BRSA, BTUP, EGD, IRAD, QGTR, RGBM, UGBM and UPGD could be relevant for T1CE contrast agent uptake analysis. Model training for tumor segmentation could benefit from the BGBM, BTUP, IRAD, RGBM, UGBM and UPDG datasets.

As part of our ongoing work, we are developing tools that facilitate easier and standardized access to these datasets within the research community. The search for datasets for this study stopped in May 2024, and new datasets are being released regularly. A natural continuity of this work would involve tracking the emergence of new datasets and updating this review accordingly. More generally, it is important for medical and research communities to collaborate in the creation of such datasets in order to respect as much as possible the quality criteria expected from both fields - imaging data alone is not sufficient for many research questions, but coupling with molecular information is needed. An ideal scenario would be publishing such datasets while finding a mechanism that leaves the door open for updates in case of changes such as the WHO classification. Another idea would be to agree on minimum specific publication quality criteria for such datasets to reach a standard dataset format in the future.      

\section*{Acknowledgements}
AH, RM and FL acknowledge support from the Luxembourg National Research Fund (FNR) through grant INTER/DFG/21/15020234 which is co-funded by the Deutsche Forschungsgemeinschaft (DFG) project number 458610525. 
FL is a fellow under a donation of Quilvest S.A. MAA is a fellow of the Institute of Advanced Studies of the University of Luxembourg.  
Data storage for dataset evaluation presented in this report was carried out
using the HPC facilities of the University of Luxembourg~\cite{VBCG_HPCS14}.
The authors thank Beatriz Garcia Santa Cruz for discussions.

\section*{Author Contributions}
MAA Data curation, Investigation, Validation, Methodology, Visualization, Writing - original draft; RM Software, Methodology; FH Conceptualization, Writing - review \& editing; FL Methodology, Investigation, Supervision, Validation, Writing - original draft; AH Conceptualization, Funding acquisition, Methodology, Project administration, Supervision, Writing - review \& editing

\clearpage
\newpage
\clearpage
\bibliographystyle{unsrt}
\bibliography{BM.bib}


\end{document}